\pdfoutput=1
\RequirePackage{fix-cm}
\documentclass[twocolumn]{svjour3}          
\smartqed  
\usepackage[numbers]{natbib}

\usepackage{enumerate}
\usepackage{algorithm2e}
\usepackage{booktabs}
\usepackage{multirow}
\usepackage{caption}
\usepackage{subcaption}
\usepackage{enumitem}
\usepackage{tabularx}
\usepackage{graphicx}
\usepackage{color}
\usepackage{amsmath}
\usepackage{fancyvrb}

\spnewtheorem{nlq}{NL Query}{\bfseries}{\itshape}
\spnewtheorem{tsql}{Translated SQL}{\bfseries}{\rmfamily}

\begin{document}

\title{xDBTagger: Explainable Natural Language Interface to Databases Using Keyword Mappings and Schema Graph
}


\author{Arif Usta         \and
        Akifhan Karakayali \and
        \"{O}zg\"{u}r Ulusoy
}


\institute{Arif Usta \at
              University of Waterloo \\
              Waterloo, ON, Canada\\
              \email{arif.usta@uwaterloo.ca}           
           \and
           Akifhan Karakayali \at
            The Central Bank of the Republic of T\"{u}rkiye\\
            Ankara, Turkey \\
            \email{akifhan.karakayali@tcmb.gov.tr}
            \and
            \"{O}zg\"{u}r Ulusoy \at
            Bilkent University \\
            Ankara, Turkey \\
            \email{oulusoy@cs.bilkent.edu.tr}
}

\date{Received: date / Accepted: date}

\maketitle

\begin{abstract}
Translating natural language queries (NLQ) into structured query language (SQL) in interfaces to relational databases is a challenging task that has been widely studied by researchers from both the database and natural language processing communities. Numerous works have been proposed to attack the natural language interfaces to databases (NLIDB) problem either as a conventional pipeline-based or an end-to-end deep-learning-based solution. Nevertheless, regardless of the approach preferred, such solutions exhibit black-box nature, which makes it difficult for potential users targeted by these systems to comprehend the decisions made to produce the translated SQL. To this end, we propose xDBTagger, an explainable hybrid translation pipeline that explains the decisions made along the way to the user both textually and visually. We also evaluate xDBTagger quantitatively in three real-world relational databases. The evaluation results indicate that in addition to being fully interpretable, xDBTagger is effective in terms of accuracy and translates the queries more efficiently compared to other state-of-the-art pipeline-based systems up to $10000$ times.
\end{abstract}

\section{Introduction}
\label{intro}


SQL is used as a standard tool to extract data out of a relational database. Although SQL is a powerfully expressive language, even technically skilled users have difficulties using SQL. Along with the syntax of SQL, one has to know the schema underlying the database upon which the query is issued, which further causes hurdles in using SQL. Consequently, casual users find it even more challenging to express their information needs, which makes SQL less desirable. To remove this barrier, an ideal solution is to provide a search engine-like interface in databases. The goal of \textit{NLIDB} is to break through these barriers to make it possible for casual users to employ their natural language to extract information.

Recently, many works have been developed attacking the NLIDB problem; such as conventional pipeline-based approaches \cite{SODA, NALIR, ATHENA, Sqlizer, athena++} or end-to-end deep-learning-based approaches \cite{iyer-etal-2017-learning, zhong2017seq2sql, xu2017sqlnet, syntaxSQL, IRNet2019-towards, dbpal2020, lin-etal-2020-bridging, scholak-etal-2021-picard}. Neural network-based solutions seem promising in terms of translation accuracy and robustness, covering semantic variations of queries. However, they struggle with queries requiring translation of complex SQL queries, such as aggregation and nested queries, especially if they include multiple tables. They also have a huge drawback in that they need many SQL-NL pairs for training to perform well, which makes pipeline-based or hybrid solutions still an attractive alternative. \cite{challenges2020}.

Whether it is a pipeline-based or an end-to-end deep learning approach, existing solutions have black-box nature when it comes to outputting translated SQL. Being a black-box solution makes it difficult for users to understand how the result SQL is produced along the way, which is a vital defectiveness for any modern intelligent system that should aim to gain the trust of the users \cite{KBSystemsExp}. This undesirable property of NLIDB solutions becomes much more consequential in an online scenario, especially for casual users (i.e., users with little to no technical expertise in SQL), who are the primary potential audience targeted by the problem of NLIDB.

Although NLIDB is a well-studied problem in the literature, the transparency and explainability of the proposed solutions have been overlooked. An intelligent system such as an NLIDB system has to be transparent and self-explanatory to the user so that they can comprehend the decisions made by the system. As highlighted by many previous studies \cite{KBSystemsExp, gregor1999explanations, poulin2006visual_exp, gunning2019darpa}, having an explainable intelligent system exhibits many benefits including but not limited to improving users' trust in the system, helping users understand the decisions made by the system,  and showing the limitations of the systems for certain use-cases, all of which can be instrumental towards developing more user friendly and preferable NLIDB systems. 


Consider the below pair of NL query and SQL translation from a movie database domain to  understand better how an explainable NLIDB solution would be handy and, in fact, essential for users to reason with the results:

\begin{nlq}
\textit{Who is the director of the series House of Cards produced by Netflix?}
\end{nlq}

\begin{tsql}
    \begin{verbatim}
    Select * From
    tv_series, copyright, company, 
    directed_by, director
    Where (tv_series.msid = copyright.msid)
    and (copyright.cid = company.cid)
    and (tv_series.msid = directed_by.msid)
    and (directed_by.did = director.did)
    and (tv_series.title = "House of Cards")
    and (company.name = "Netflix")
    \end{verbatim}
\end{tsql}

Although such a SQL query is easy to understand for an expert user at a glance, it is difficult to interpret for casual users, which are the ones targeted by the NLIDB problem. In the above query, there are 2 keywords, \textit{House of Cards} and \textit{Netflix}, found under the attributes \textit{title} and \textit{name} of their respective entity tables \textit{tv\_series} and \textit{company}. However, to find the corresponding tuple(s) matching these two values from different tables, SQL requires a join operation; in this case, a 5-way join. The subsequent four lines after the \textit{Where} clause in the above example represent conditions to ensure the right join. More importantly, although one needs access to 3 entity tables; 2 (tv\_series and company) for utterances found in the query and 1 (director) for the desired information asked by the user, SQL requires two more intermediate tables, \textit{copyright} and \textit{directed\_by}, to complete the join. In an ideal NLIDB, the story behind the translated SQL, such as above, should be provided to the user to some extent. In addition to the explanations needed for understanding SQL structure, the NLIDB system should also ideally give explanations for how it matches schema elements (e.g., three tables and two attributes for the above example) to the corresponding utterances.

To address above-mentioned concerns, we propose an explainable, end-to-end NLIDB solution, \textbf{Explainable DBTagger (xDBTagger)}, by extending our previous work in \cite{dbtagger}. xDBTagger is a hybrid solution utilizing both deep learning and rule-based approaches. To the best of our knowledge, xDBTagger is the first study exercising explainable artificial intelligence (XAI) paradigm in the NLIDB problem. In what follows, we list the main contributions of our work:



\begin{itemize}[leftmargin=*,topsep=0pt]
    \item We use our previous work \cite{dbtagger}, which is a deep learning model specifically tailored for sequence tagging in NLIDB, to extract keyword mappings given the NLQ.  
    \item We propose a novel wrapper tailored for sequence tagging problems around a state-of-the-art XAI work, LIME \cite{lime}, to explain decisions made for keyword mappings output for each token in NLQ. We provide explanations for each keyword mapping corresponding to tokens in NLQ by highlighting both the positive and negative contributions of each surrounding token. 
    \item We propose an effective and efficient SQL translation algorithm suitable for interpretability by utilizing keyword mappings and schema graphs. We provide textual and visual explanations for the user to comprehend how the translation algorithm works. 
    \item We quantitatively evaluate the entire pipeline in three publicly available datasets where xDBTagger performs competitively as the most efficient and scalable solution.
    \item We deploy xDBTagger in a user-friendly and interpretable interface in which the user is presented the translated SQL along with the explanations for the decisions made throughout the pipeline. 
\end{itemize}

\begin{figure*}[!t]
  \includegraphics[width=1.0\textwidth]{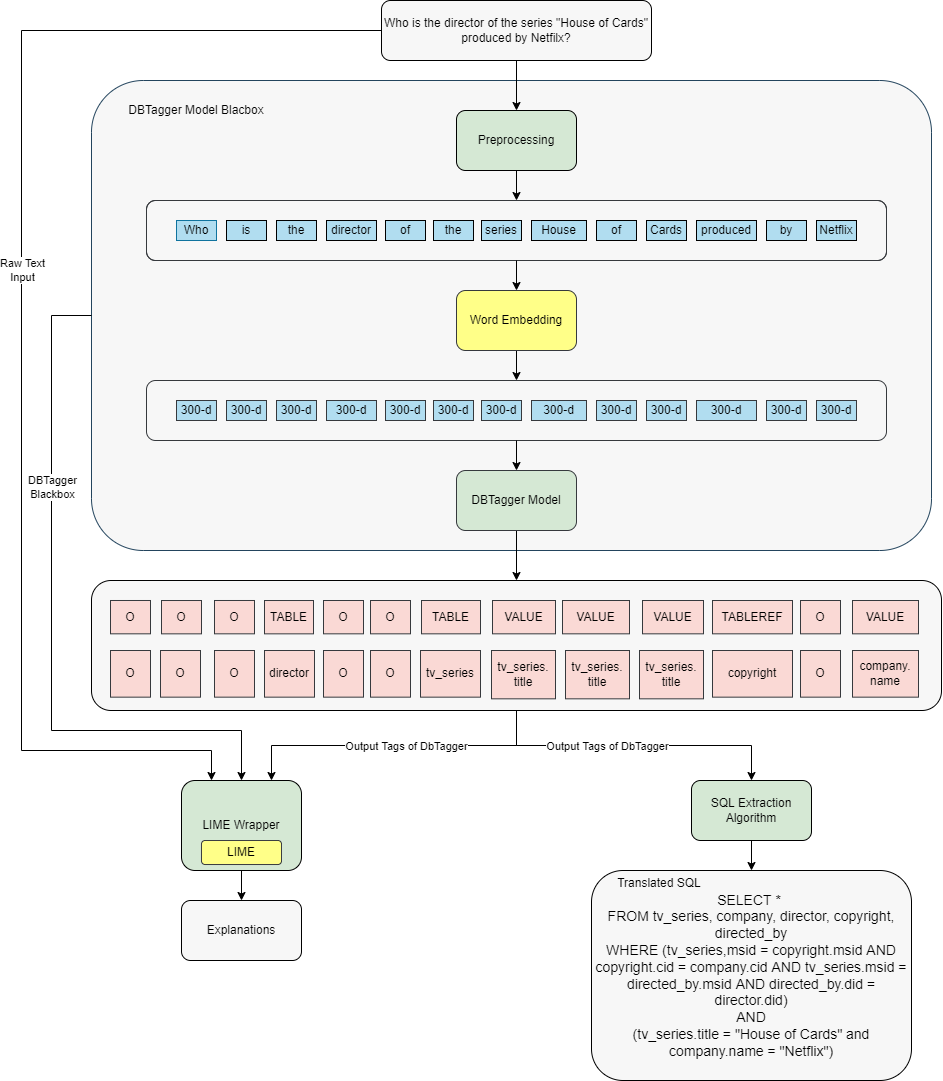}
\caption{System Architecture of xDBTagger}
\label{fig:flowchart}       
\end{figure*}

The remainder of the paper is organized as follows. In the next section, we give an overview of the system architecture of xDBTagger. Section \ref{sec:keywordMapper} presents the neural network structure we design for the keyword mapping step. We explain how we modify LIME to produce explanations for keyword mappings output by DBTagger in Section \ref{sec:lime-discussion}. We thoroughly review the main components of the SQL extraction algorithm in Section \ref{sec:SQLExtract}. 
In Section \ref{sec:experiments}, we provide quantitative experimental results for both the keyword mapper and the entire SQL translation pipeline (Section \ref{subsec:keywordMappingResults}). In addition to quantitative results, we illustrate the user interface and provide examples of textual and visual explanations in the interface (Section \ref{sec:user-interface}). We summarize the related work and conclude the paper in Sections \ref{sec:relatedWork} and \ref{sec:conclusion}, respectively.

\section{System Architecture}

\begin{figure*}[t]
    \centering
    \includegraphics[width=\textwidth, height=6cm,keepaspectratio]{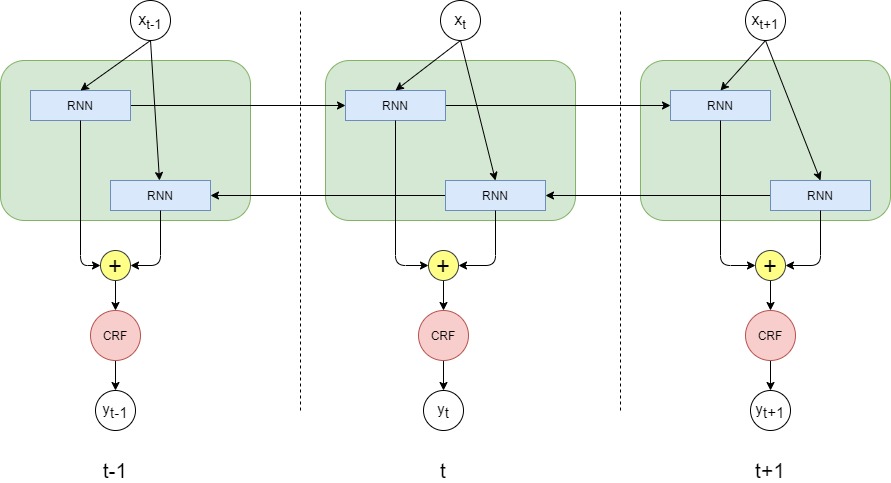}
    \caption{Deep Sequence Tagger Network}
    \label{fig:NNStructure1}
\end{figure*}

Figure \ref{fig:flowchart} depicts an overview of the translation pipeline along with explanation components to make the decision-making throughout the pipeline interpretable. The workflow starts with an input NLQ from the user. The query first goes through pre-processing, which removes special characters and punctuations such as commas and quotes. After removing those characters, the query is tokenized into words using spaces. These tokens are then converted to 300-dimensional vector representations using a pre-trained word embedding model. In our implementation, we used a pre-trained fast-text~\cite{fasttext} model. 

The output of the embedding model $X= [x_1, x_2, ... , x_n]$ is an array of 300-dimensional vectors where $n$ is the length of the query (i.e., the number of tokens in the NLQ). Each 300-dimensional vector is a representation of each word in the original query. Following that, $X$ is fed as the input to DBTagger model, which outputs corresponding keyword mappings for each token in the query. DBTagger outputs 2 series of outputs; 1 for type tags (i.e., schema element such as table, attribute if the token is relevant for SQL translation or "O" if irrelevant) and 1 for schema tags (i.e., deeper level tags such as name of a table or an attribute) of the tokens in the NLQ. 

We use LIME \cite{lime} to explain keyword mappings output by DBTagger. LIME requires a black-box model which can output prediction probabilities and raw input text for explanation. To satisfy these conditions, we construct a DBTagger Model Blackbox, which takes the raw text as input and gives predictions with probabilities as output. 
Vanilla LIME tries to explain a single classified output given a sequential input text. However, it is not the case in our problem (i.e., sequence tagging problem requiring a classification and, therefore, an explanation for each token in the query). To alleviate this issue, we implement a wrapper around LIME which is explained in Section \ref{sec:taglime}. This wrapper takes DBTagger Model Blackbox, raw text input, and predicted output tags to produce an explanation for each token.

\section{Keyword Mapper - DBTagger}
\label{sec:keywordMapper}

\begin{figure*}[t!]
    \centering
    \includegraphics[width=\textwidth, height=12cm,keepaspectratio]{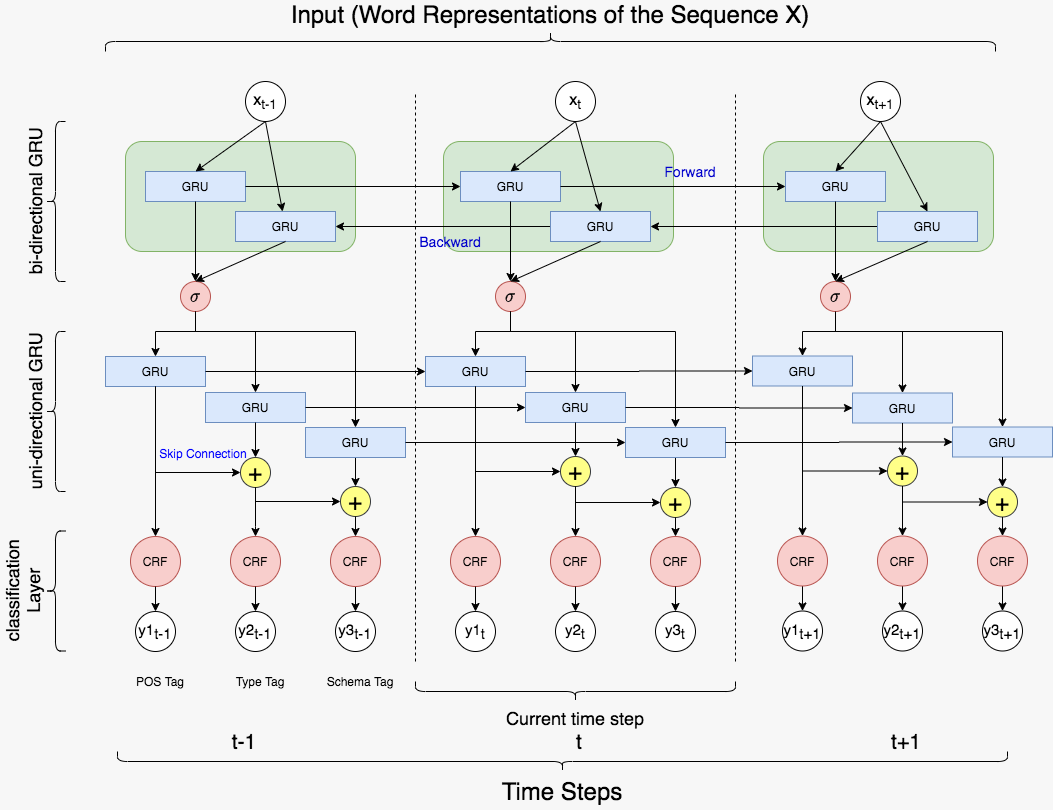}
    \caption{DBTagger Network}
    \label{fig:NNStructure2}
\end{figure*}

In this section, we first provide background information about the neural network structure utilized for sequence tagging problems such as POS tagging and NER in the NLP community. Next, we explain the network structure of \textit{DBTagger}, our keyword mapper solution in the pipeline, by pointing out modifications we introduce on top of the state-of-the-art sequence tagging architecture. Lastly, we discuss how we annotate three different class labels of tokens to employ multi-task training.

\subsection{Deep Sequence Tagger Architecture}
POS tagging and NER refer to sequence tagging problems in NLP for a particular sentence to identify parts-of-speech such as noun, verb, and adjective and locate any entity names such as person and organization, respectively. We argue that these problems are formally similar to the keyword mapping problem in NLIDB.
Recurrent Neural Networks (RNN) are at the core of architectures to handle such problems since they are a family of networks that perform well on sequential data input such as a sentence. In this particular problem, sequence tagging (\textit{keyword mapping}), RNNs are employed to output a sequence of labels for the original sentence (the query), input as a sequence of words. 

In RNN networks, the basic goal is to carry past information (previous words) to future time steps (future words) to determine values of inner states and, consequently, the final output, which makes them preferable architecture for sequential data. Given $x_t$ as input at time step $t$, calculation of hidden state $h_t$ at time step $t$ is as follows:

\begin{equation}
    h_t = f(Ux_t + Wh_{t-1})
\end{equation}

In practice, however, RNN networks suffer from \textit{vanishing gradient problem}; therefore, the limitation was overcome by modifying the gated units of RNNs; such as LSTM \cite{LSTM} and GRU\cite{GRU}. Compared to vanilla RNN, LSTM has \textit{forget gates} and GRU comprises of \textit{reset} and \textit{update} gates additionally. 
We experimented with both structures and chose GRU for its better performance in our experiments. In GRU, Update Gates decide what information to throw away and what new information to add, whereas Reset Gate is utilized to decide how much past information to forget. The calculation of GRU is as follows:

\begin{align}
    z &= \sigma(U_z.x_t + W_z.h_{t-1}) \\
    r &= \sigma(U_r.x_t + W_r.h_{t-1}) \\
    z_t &= tanh(U_z.x_t + W_s.(h_{t-1}\bullet r)) \\
    z_t &= \sigma(U_zx_t + W_zh_{t-1})
    \label{eq:2}
\end{align}

In the sequence tagging problem, in addition to past information, we also have future information at a given specific time, $t$. For a particular word $w_i$, we know the preceding words (past information) and succeeding words (future information), which can be further exploited in the particular network architecture called, \textit{bi-directional RNN} introduced in \cite{bi-directionalRNN}. Bi-directional RNN has two sets of networks with different parameters called forward and backward. The concatenation of the two networks is then fed into the last layer, where the output is determined. This process is demonstrated in Figure \ref{fig:NNStructure1}.

Sequence tagging is a supervised classification problem where the model tries to predict the most probable label from the output space. For that purpose, although conventional \textit{softmax} classification can be used, \textit{conditional random field (CRF)} \cite{CRF} is preferred. Unlike independent classification by softmax, CRF tries to predict labels sentence-wise by considering labels of the neighboring words as well. This feature of CRF is what makes it an attractive choice, especially in a problem like \textit{keyword mapping}. 
This finding was also reported in \cite{lample-etal-2016-neural}, where authors claim that CRF as the output layer gives $1.79$ more accuracy compared to the softmax layer in NER task. The final outlook of the architecture of deep sequence tagger is depicted in Figure \ref{fig:NNStructure1}.

\subsection{DBTagger Architecture}

Formally, for a given NL query, input $X$ becomes a series of vectors $[x_1,x_2,...x_n]$ where $x_i$ represents the $i^{th}$ word in the query. Similarly, output vector $Y$ becomes $[y_1,y_2,...y_n]$ where $y_i$ represents the label (actual tag) of the $y^{th}$ word in the query. Input must be in numerical format, which implies that a numerical representation of words is needed. For that purpose, the word embedding approach is state-of-the-art in various sequence tagging tasks in NLP \cite{collobert2011} before feeding into the network. So, the embedding matrix is extracted for the given query, $W\in R^{nxd} $, where $n$ is the number of words in the query and $d$ is the dimension of the embedding vector for each word. 
For the pre-calculated embeddings, we used fastText\cite{fasttext} due to it being one of the representation techniques considering sub-word (character n-grams) as well to deal with the out-of-vocabulary token problem better.

We consider $G$ to be 2-dimensional scores of output by the uni-directional GRU with size $n\times k$ where $k$ represents the total number of tags. $G_{i,j}$ refers to score of the $j^{th}$ tag for the $i^{th}$ word. For a sequence $Y$ and given input $X$, we define tag scores as;

\begin{equation}
    s(X,Y) = \sum_{i=1}^{n}A_{y_i,y_{i+1}} + \sum_{i=1}^{n}G_{i,y_i}
\end{equation}

where $A$ is a transition matrix in which $A_{i,j}$ represents the score of a transition from the $i^{th}$ tag to the $j^{th}$ tag. After finding scores, we define the probability of the sequence $Y$:

\begin{equation}
    p(Y|X) = \frac{e^{s(X,Y)}} { \sum_{\bar{Y}\in Y_x}^{}e^{s(X,\bar{Y})}}
\end{equation}

where $\bar{Y}$ refers to any possible tag sequence. During training, we maximize the log-probability of the correct tag sequence, and for the inference, we simply select the tag sequence with the maximum score.

In our architecture, we utilize \textit{Multi-task learning} by introducing two other related tasks; POS and type levels (shown in Figure \ref{fig:NNStructure2}). The reason we apply multi-task learning is to try to exploit the observation that actual database tags of the tokens in the query are related to POS tags. Besides, multi-task learning helps to increase model accuracy and efficiency by making more generalized models with the help of shared representations between tasks \cite{caruana1997multitask}. POS and Type tasks are trained with schema task to improve the accuracy of schema (final) tags.  For each task, we define the same loss function described above. During back-propagation, we simply combine the losses as follows;

\begin{equation}
    \begin{split}
    L_{total} &= \sum_{i=1}^{3}w_i\times L_i \textrm{ subject to}\\
    \sum_{i=1}^{3}w_i &= 1 
\end{split}
\end{equation}

where $w_i$ represents the weight of $i^{th}$ task and $L_i$ represents the loss calculated for the $i^{th}$ task similarly.

\subsection{Annotation Scheme}

We tackle keyword mapping as a sequence tagging problem, which is a supervised classification problem.   In our problem formulation, every token (i.e., words in the natural language query) associates three different tags: part-of-speech (POS) tag, type tag, and schema tag. In the following subsections, we explain how we extract or annotate each of them in detail.
\subsubsection{POS Tags}
To obtain the POS tags of our natural language queries, we used the toolkit of Stanford Natural Language Processing Group named Stanford CoreNLP\cite{coreNLP}. We use them as they are output from the toolkit without doing any further processing since the reported accuracy for POS Tagger (97\%) is sufficient enough.

\subsubsection{Type Tags}

\begin{table}[t]
\centering
\begin{tabular}{@{}ccc@{}}
\toprule

NLQ      & \begin{tabular}[c]{@{}c@{}}Type\\ Tag\end{tabular} & \begin{tabular}[c]{@{}c@{}}Schema\\ Tag\end{tabular} \\ \midrule
Who & O  & O \\

is & O  & O  \\
the & O & O \\

director & TABLE & director \\
of & O & O \\

the & O & O \\
series & TABLE & tv\_series \\

House & VALUE & tv\_series.title \\
of & VALUE & tv\_series.title \\

Cards & VALUE & tv\_series.title \\
produced & TABLEREF & copyright \\

by & O & O \\
Netflix & VALUE & company.name \\ \bottomrule
\end{tabular}
\caption{An example NL query with its tags corresponding to each word in two target levels}
\label{tab:dbtagger-predictions}
\end{table}

In each natural language query, there are keywords (words or consecutive words) which can be mapped to database schema elements such as table, attribute, or value. We divide this mapping into two levels; type tagging and schema tagging. Type tags represent the type of the mapped schema element to be used in the SQL query. In total, we have seven different type tags;
\begin{itemize}[leftmargin=*]
    \item \textbf{TABLE}: NLQs contain nouns that may inhibit direct references to the tables in the schema, and we tag such nouns with \textit{TABLE} tag. In the example NL query given in Figure \ref{fig:flowchart}, noun \textit{director} has a type tag as TABLE, which also supports the intuition that schema labels and pos tags are related.
    \item \textbf{TABLEREF}: Although the primary sources for table references are nouns, some verbs contain references to the tables, most of which are relation tables. TABLEREF tag is used to identify such verbs. Revisiting the example given in Figure \ref{fig:flowchart}, the verb \textit{produced} refers to the table \textit{copyright}, and therefore it is tagged with TABLEREF to differentiate better the roles of POS tags in the query.
    \item \textbf{ATTR}: In SQL queries, attributes are mostly used in SELECT, WHERE, and GROUP BY clauses. Natural language queries may contain nouns that can be mapped to those attributes. We use the ATTR tag for tagging such nouns in natural language queries.
    \item \textbf{ATTRREF}: Like the TABLEREF tag, the ATTRREF tag is used to tag the verbs in the natural language query that can be mapped to the attributes in the SQL query.
    \item \textbf{VALUE}: In NLQs, there are many entity-like keywords that need to be mapped to their corresponding database values. These words are mostly tagged as \textit{Proper noun-NNP} such as the keyword \textit{House of Cards} in the example query. In addition to these tags, it is also likely for a word to have a \textit{noun-NN} POS tag with a \textit{Value} tag corresponding to schema level. In order to handle these cases having different POS tags, we have \textit{Value} type tags (e.g., \textit{House} keyword in the example query is part of a keyword that needs to be mapped as \textit{value} to \textit{tv\_series.title}). Keywords with \textit{Value} tags can later be used in the translation to determine "where" clauses in SQL.
    \item \textbf{COND}: After determining which keywords in the query are to be mapped as values, it is also important to identify the words that imply which type of conditions to be met for the SQL query. For that purpose, we have the \textit{COND} type tag.
    \item \textbf{O (OTHER)}: This type of tag represents words in the query that are not needed to be mapped to any schema instrument related to the translation step. Most stop words in the query (e.g., the) fall into this category. 
\end{itemize}


\subsubsection{Schema Tags}
Schema tags of keywords represent the database mapping that the keyword is referring to, the name of a table, or the attribute. Tagging a keyword with a type tag is important yet incomplete. To find the exact mapping the keyword refers to, we define a second-level tagging where the output is the name of the tables or attributes. For each entity table (e.g., \textit{movie} table in the shortest path component of Figure \ref{fig:nlidbinterface}) and for each non-PK or non-FK attribute (attributes which have semantics) we define a schema tag (e.g., \textit{movie, people, movie.title}, etc., referring to Figure \ref{fig:nlidbinterface}). We complete possible schema tags by carrying \textit{OTHER} and \textit{COND} from type tags. 
We use the same schema tag for attributes and values (e.g., \textit{movie.title}), but differentiate them at the inference step by combining tags from both type tags and schema tags. If a word is mapped into \textit{Value} type tag as a result of the model, its schema tag refers to the attribute in which the value resides.

In order to annotate queries, we annotate each word in the query for three different levels mentioned above. While POS tags are extracted automatically, we manually annotate the other two levels. Annotations were done by three graduate and three undergraduate computer science students who are familiar with database subject. Although annotation time varies depending on the person, on the average, it took a week to annotate tokens by a single person for two levels (type and schema) for a query log with $150$ NL questions, which we believe is practical to apply in many domains.

\section{Explanations for Keyword Mapper}
\label{sec:lime-discussion}

In this section, we provide the details about the techniques we employ to explain the decisions made by DBTagger, our keyword mapper in the pipeline. First, we give a short overview of the LIME \cite{lime} work and highlight its applicability and limitation in the context of the sequence tagging problem. Next, we explain how we tailor LIME to the sequence tagging problem (i.e., classification problem for each item in the sequence) to deploy in our pipeline. 

\subsection{LIME}
\label{sec:taglime}


LIME \cite{lime} is short for "Local Interpretable Model-Agnostic Explanations", where each part in the name exhibits a desirable property a black-box explanation model must have. "Local" implies that LIME is an outcome explanation model, explaining the decision made on a particular instance, which is in line with our goal. "Model-agnostic" refers that LIME works with any type of input data (e.g., image, text) or a black-box model (e.g, a linear classifier such as logistics regression or a neural network based model), which is one of the reasons why we use LIME in xDBTAgger that includes a neural network based keyword mapper that we want to explain to the user. 

Interpretation and explanation are important terms often used interchangeably in the context of XAI; however, they have distinct meanings. The former is more involved in providing abstracts in a way humans can make sense of, whereas the latter revolves around highlighting important features that play a role in decision-making for a given instance \cite{xaiSurvey1}. Analogously, the explanations that are not interpretable are useless, which is addressed by LIME. LIME argues that interpretable data representations differ from actual feature representations by asserting that interpretable data representations, such as binary vectors stating the existence of a word, are easily understood by humans. In contrast, actual feature representations, such as word embedding vectors, are not that straightforward and comprehensible. This distinctness is crucial since explanations produced by LIME are based on interpretable data representations. 

In particular, LIME provides the importance of the features for a given instance as explanations, and to make them interpretable it follows a binary approach that highlights how important certain parts of the input are when they are present or absent. For a given input, LIME perturbs the input by randomly removing parts of the input and tries to understand how the model behavior changes. For instance, LIME creates a series of artificial sentences for a particular text input as a sentence in which random tokens are removed. LIME then tries to assign an importance score to each token for the decision (e.g., a target label by a classifier) by weighing the changes in model behavior. If the score is positive, the token is helpful when deciding the outcome for a particular input, whereas it is disadvantageous to the outcome when the score is negative. The absolute value of the score implies the contribution the token makes to the outcome, either positively or negatively.

\subsection{LIME Wrapper}

Due to its properties, LIME is applicable in classification problems where the input is a sentence, and it is important to explain the importance of each token in the sentence in deciding a particular class. Also, note that the architecture of DBTagger, our keyword mapper, also utilizes signals from neighboring tokens when deciding the type and schema classes of a particular token by using CRF (see Figure~\ref{fig:NNStructure2}) at the last layer. This property of DBTagger aligns perfectly with the applicability of LIME in a text classification problem, as explained above. However, vanilla LIME is not directly applicable where the model generates multiple outcomes for a given sentence. In other words, vanilla LIME produces explanations for models that classify the whole text sequence into one class (e.g., sequence classification such as sentiment analysis), whereas in our case, there is a classification for every token in a sentence, referred to as sequence tagging problem in NLP. Hence, we make modifications and add a wrapper around LIME to output explanations for each token suitable for DBTagger. 


In particular, the wrapper around LIME uses four groups of information; the NLQ (i.e., the text input as a list of tokens), DBTagger Model Black-box (i.e., the probabilities of target classes for each token), predicted type and schema classes of DBTagger Model, and output mask to perturb the sentence suitable for keyword mapping problem. The main purpose of this wrapper is to coordinate the communication between LIME and the output mask. With the help of the output mask, it becomes possible for LIME to produce an explanation for a specific token; however, we still need to select the token that will be explained. To achieve this, we only explain the tokens with a predicted tag other than $"O"$ for time efficiency since LIME uses many resources, and producing explanations is a time-consuming process, even for a single token. Once the token selection is made, the wrapper adjusts the output mask so that the model gives the output for the selected token. Therefore LIME can analyze the output and produce an explanation for that token. This process is repeated for every token that is selected for the explanation.

\section{SQL Translation Algorithm}
\label{sec:SQLExtract}

We use simple yet effective algorithms to construct the translated query given a set of type and schema tags output by the keyword mapper, DBTagger. In particular, for the translation algorithm, we have three channels of input (see Figure \ref{fig:flowchart}); (i) type tags (i.e., lists of tags indicating whether each word in NLQ is a table, column or value) and (ii) schema tags (i.e., names of the schema elements found such as table or column names for each word having a valid type tag) output by DBTagger, and (iii) the input query.

Utilizing the above-mentioned inputs, the SQL translation algorithm has 4 main components, which are (i) \textit{Schema Graph Extraction}, (ii) \textit{Join-Path Inference}, (iii) \textit{Where Clause Completion} and (iv) \textit{Heuristics for Aggregate Queries}. Each component is explained in detail in the following subsections.

\begin{algorithm}[t]
  \SetAlgoLined
  \caption{Generating the schema graph out of the database}
  \SetKwInOut{Input}{Inputs}
  \SetKwInOut{Output}{Outputs}
  \SetKwProg{ExtractGraph}{ExtractGraph}{}{}
  
   \ExtractGraph{$(D)$}{
    \Input{Relational database $D$}
    \Output{Graph model of database $dbGraph$}
     $dbGraph \gets Graph()$\;
        \ForEach{table $t_i \in D$}{%
            $dbGraph.addNode(t_i)$
            
            \ForEach{column $c_i \in t_i$}{
                \uIf{$s_i \notin dbGraph$}{
                    $dbGraph.addNode(c_i)$
                }
                $dbGraph.addEdge(t_i,c_i)$
            }
        }
        \KwRet{$dbGraph$}\;
   }
   \label{algo1}
\end{algorithm}

\subsection{Schema Graph Extraction}
\label{subsec:SQLGraphExtract}

As the first step of the translation algorithm, we construct a \textit{schema graph} out of the database upon which the NLQ is issued. In the graph, each node represents either a table or an attribute in the database. Each connection between a pair of nodes in the graph is an undirected edge, connecting table and attribute nodes to represent their \textit{has-A} relations. An edge in the schema graph can only be between two different types of nodes (i.e., between a table and an attribute).

\begin{figure}[t]
  \includegraphics[width=0.5\textwidth]{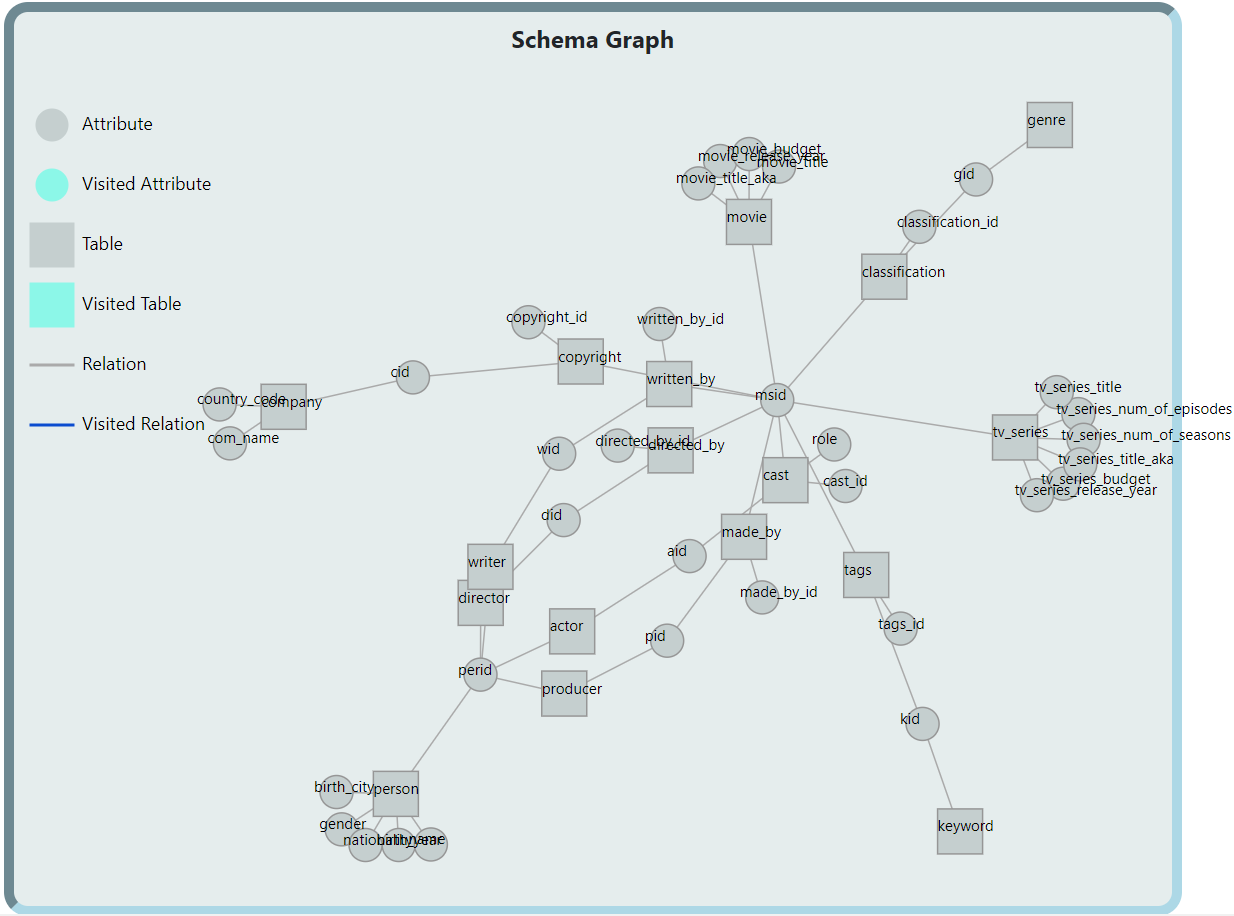} 
\caption{An example schema graph consisting of tables, columns and their relations}
\label{fig:labelledjoinpath}       
\end{figure}

Algorithm \ref{algo1} shows how to extract a simple schema graph from a relational database assuming that foreign keys and primary keys have the same name. This assumption can be relaxed by adding edges between foreign keys and primary keys if the given database schema contains tables having different name primary-foreign pairs, or multiple foreign keys referencing the same primary key, or self references. An example schema graph extracted from the type and schema tags output by DBTagger for the example NLQ given in Section 1 is depicted in Figure \ref{fig:labelledjoinpath}.

\begin{algorithm}[t]
  \SetKwBlock{Begin}{Begin}{}
  \SetAlgoLined
  \caption{Inferring Shortest Join-Path}
  \SetKwInOut{Input}{Input}
  \SetKwInOut{Output}{Output}
  \SetKwProg{ExtractJoinRelation}{ExtractJoinRelation}{}{}
  
   \ExtractJoinRelation{$(G, T)$}{
    \Input{Database Graph $G$ and list of tables $T$}
    \Output{Graph paths that contains SQL join information}
    $joinPath \gets \emptyset$\;
    $candidate \gets null$\;
        \ForEach{table $t_i \in T$}{%
            \ForEach{table $t_j \in T$}{
                \uIf{$t_i \neq t_j$}{
                    $paths \gets findShortestPaths(G, t_i, t_j)$
                    
                    \ForEach{path $p \in paths$}{
                        \uIf{$T \subseteq p$}{
                            $returnPaths.append(p)$\;
                            \KwRet{$returnPaths$}\;
                        
                        }
                        \uElse{
                            $missingTables \gets T \setminus p$\;
                            \uIf{$candidate == null$}{
                                $candidate \gets $ $(p, missingTables)$\;
                            }
                            \uElse {
                                \uIf{$length(missingTables) < length(candidate_1)$}{
                                    $candidate \gets $ $(p, missingTables)$\;
                                }
                            }
                            
                        }
                    }
                }
            }
        }
        $returnPaths.append(candidate_0)$\;
        \ForEach{table $t \in candidate_1$}{
            $paths \gets findShortestPaths(G, candidate_{00}, t)$\;
            \ForEach{$path \in paths$}{
                $listReduced \gets False$\;
                \ForEach{table $t_2 \in candidate_1$}{
                    \uIf{$t_2 \neq t \land t_2 \in path$}{
                        $candidate_1.remove(t_2)$\;
                        $listReduced \gets True$\;
                    }
                }
                \uIf{$listReduced$}{
                    $returnPaths.append(path)$\;
                }
            }
        }
        \KwRet{$returnPaths$}\;
   }
   \label{algo2}
\end{algorithm}

\subsection{Join-Path Inference}
\label{subsec:sqlJoinPath}

After constructing the schema graph, we try to find the shortest way of combining all the nodes representing schema elements (i.e., tables and attributes). Recall that we have type and schema tags output by DBTagger to work with. We use Dijkstra's shortest path algorithm to find the in-between nodes required to infer the join-path.

The entire algorithm is given in Algorithm \ref{algo2}. Using the type and schema tags output by DBTagger, a set of tables $T$ is created containing all the related tables for the given NLQ. $T$ is composed of found tables (i.e., name of the tables for those words that have \textit{TABLE} or \textit{TABLEREF} as type tag), and tables of found attributes (i.e., name of the columns for those words that have \textit{ATTR} or \textit{ATTRREF} or \textit{VALUE} as type tag).
After that, $T$ is given as an input to Algorithm \ref{algo2} with the schema graph, and the shortest paths that contain and join the tables in $T$ are found. An example join-path found for the given NLQ is depicted in Figure \ref{fig:labelledjoinpath}. In the path, each consecutive nodes connected through an attribute node exhibits a join condition (e.g., \textit{director} and \textit{directed\_by} tables require a join through attribute \textit{did}).


\subsection{Where Clause Completion}
\label{subsec:sqlWhereClause}

The next step is to construct WHERE conditions of SQL. The process starts with gathering the outputs of type and schema tags from DBTagger. A two dimensional array $M$ is created where $M_{i,1}$ contains the $i^{th}$ query token, $M_{i,2}$ contains type tag of the $i^{th}$ token, and $M_{i,3}$ contains the schema tag of the $i^{th}$ token. Following that, a pre-processing step is applied on the array $M$ to smooth out consecutive tokens that have the same mapping information by merging them together. In particular, the tokens that have \textit{VALUE} as a type tag inhibit WHERE conditions to focus. For each list of consecutive tokens with $VALUE$ mapping, a pair of the query token (e.g., House of Cards) and column name of the respective table (e.g., title column of the tv\_series table) is created and added to where conditions of SQL. Algorithm \ref{algo3} shows how WHERE conditions are extracted from the given list $M$.

%
\begin{algorithm}[t]
  \SetAlgoLined
  \caption{Extraction of SQL WHERE Conditions}
  \SetKwInOut{Input}{Input}
  \SetKwInOut{Output}{Output}
  \SetKwProg{ExtractWhereConditions}{ExtractWhereConditions}{}{}
  
   \ExtractWhereConditions{$(M)$}{
    \Input{Two dimensional array $M$, containing the NLQ and keyword mapping information}
    \Output{SQL WHERE conditions $whereConditions$}
     $whereConditions \gets \emptyset$\;
     $M \gets mergeConsecutiveMappings(M)$ \tcp*[1]{Merges multi-word entities to a single mapping e.g: Brad Pitt}
     
        \ForEach{token $k_{1i}, k_{2i}, k_{3i} \in M_1, M_2, M_3$}{%
            \uIf{$k_{2i}$ is VALUE}{
                $whereConditions.add(k_{1i},k_{3i})$ \tcp*[1]{$k_{1i}$ and $k_{3i}$ contains the keyword and column information of that keyword respectively}
            }
        }
        \KwRet{$whereConditions$}\;
   }
   \label{algo3}
\end{algorithm}

\begin{algorithm}[t]
  \SetAlgoLined
  \caption{Extraction of SQL AGGREGATE Clause}
  \SetKwInOut{Input}{Input}
  \SetKwInOut{Output}{Output}
  \SetKwProg{ExtractAggregateClause}{ExtractAggregateClause}{}{}
  
   \ExtractAggregateClause{$(M, prevWindow)$}{
    \Input{Two dimensional list $M$ containing natural language query and keyword mapping information, $prevWindow$ parameter for aggragete keyword search in sentence}
    \Output{SQL AGGREGATE Clause}
        $SUMKeywords \gets getSUMKeywords()$;
        $COUNTKeywords \gets getCOUNTKeywords()$;
        $AVGKeywords \gets getAVGKeywords()$;
        \ForEach{token $k_{1i}, k_{2i}, k_{3i} \in M_1, M_2, M_3$}{%
            \uIf{$k_{2i} \in [TABLE, TABLEREF, ATTR, ATTRREF]$}{
                \ForEach{token $k_{1j} \in [M_{1 i-prevWindow}, ..., M_{1i}]$}{
                    \uIf{$k_{1j} \in SUMKeywords$}{
                        \KwRet{$(SUM,k_{1i},k_{2i},k_{3i})$}\;
                    }
                    \uIf{$k_{1j} \in COUNTKeywords$}{
                        \KwRet{$(COUNT,k_{1i},k_{2i},k_{3i})$}\;
                    }
                    \uIf{$k_{1j} \in AVGKeywords$}{
                        \KwRet{$(AVG,k_{1i},k_{2i},k_{3i})$}\;
                    }
                }
            }
        }
        \KwRet{$None$}
   }
   \label{algo4}
\end{algorithm}

\subsection{Heuristics for Aggregate Queries}
\label{subsec:aggHeuristic}

Finally, we used a simple and effective technique to detect potential aggregate operations for the constructed SQL. There are some specific keywords -such as total, many, count etc.- that imply certain aggregate operations. Using these keywords, we define keyword sets for each aggregate operation and perform a search in the NLQ for potential aggregate keywords. Algorithm \ref{algo4} shows the outline of the performed search. After retrieving the mapping output from DBTagger, we select the words that have TABLE, TABLEREF, ATTR, or ATTRREF as the type tag as our candidates for aggregation. For each keyword set, we search the words that appear before our candidates. If we find a matching keyword, we return the candidate keyword, its mapping information, and the matching aggregate operation. If no matching is found, the algorithm returns $None$, implying that no aggregation should be applied. For searching the previous words of the token $k_{i}$, we define a window size $prevWindow$ and perform the keyword search for the words that are inside this window, namely $[k_{i-prevWindow}, ..., k_{i-2}, k_{i-1}]$.


\section{Experimental Setup}
\label{sec:experiments}

\subsection{Datasets}

In our experiments we used \textcolor{black}{\textit{yelp, imdb \cite{Sqlizer}}, and \textit{mas \cite{NALIR}} datasets which are heavily used in many NLIDB related works by the database community \cite{NALIR, ATHENA, Sqlizer,jagadishBridge}}. 

The statistics about each dataset for which annotation is done are shown in Table \ref{tab:comprops}. In Table \ref{tab:comprops} (referring to Figure \ref{fig:labelledjoinpath}), entity tables refer to main tables (e.g., Movie), relation tables refer to hub tables that store connections between entity tables (e.g., cast, written\_by), nonPK-FK attributes refer to attributes in any table that is neither PK nor FK (e.g., gender in People table), and finally total tags refer to a unique number of taggings extracted from that particular schema depending on the above-mentioned values. Final schema tags of a particular database are determined by composing table names and names of the nonPK-FK attributes in addition to COND and OTHER. In the last two rows of Table \ref{tab:comprops}, we show the number of annotated NL questions, referred to as queries, and the number of total words inside these queries, referred to as tokens.

\begin{table}[t]
    \centering
    \caption{\textcolor{black}{Statistics of the databases used}}
    \begin{tabular}{lccc}
    \toprule
         & \multicolumn{3}{c}{\textbf{Database}}\\
         \toprule
         Properties (\#)& imdb & mas & yelp 
         \\
         \midrule
         entity tables&6&7&2 \\
         relation tables&11&5&5 \\
         total tables&17&12&7\\
         total attributes&55&28&38 \\
         nonPK-FK attributes&14&7&16 \\
         total tags&31&19&20 \\
         queries&131&599&128 \\
         tokens in queries&1250&4483&1234 \\
         \bottomrule
    \end{tabular}
    \label{tab:comprops}
\end{table}

\subsection{Quantitative Evaluation of xDBTagger}
\label{subsec:keywordMappingResults}

\subsubsection{Keyword Mapping Evaluation}

In order to train DBTagger, the keyword mapper for the pipeline, we first split the datasets into train-validation sets with a $5-1$ ratio, respectively, to be used for tuning task weights. For models trained on multiple tasks, we used $0.1-0.2-0.7$ as tuned weights for POS, Type, and Schema tasks, respectively.

We train our \textcolor{black}{deep neural} models using the backpropagation algorithm with two different optimizers; namely Adadelta \cite{Zeiler2012ADADELTAAA} and Nadam \cite{nadam}. We start the training with Adadelta and continue with Nadam. We found that using two different optimizers resulted better in our problem. For both shared and unshared bi-directional GRUs, we use $100$ units and apply dropout \cite{dropout} with the value of $0.5$, including recurrent inner states as well. For training, the batch size is set to $32$ for all datasets. 
Parameter values chosen are similar to that reported in the study  \cite{lample-etal-2016-neural} (the state-of-the-art NER solution utilizing deep neural networks), such as the dropout and batch size values. \textcolor{black}{We measure the performance of each neural model by applying cross-validation with 6-folds. All the results reported are the average test scores of 6-folds. During inference, we discard POS and Type task results and use only Schema (final) tasks to measure scores.}

We implemented three different unsupervised approaches utilized in the state-of-the-art NLIDB works for the keyword mapping task as baselines to compare with DBTagger. We implemented sql querying over database column approaches (regex and full-text search), which is preferred in NALIR \cite{NALIR}. We implemented a well-known tf-idf baseline for exact matching by constructing an inverted index over unique database values present, as in the work ATHENA \cite{ATHENA}. We also implemented a semantic similarity matching approach in which pre-defined word embeddings are used. This approach is exercised by Sqlizer \cite{Sqlizer}. In addition to these conventional unsupervised solutions, we also implemented TaBERT \cite{tabert}, a pre-trained language model utilizing transformer architecture to compare with our proposed solution. 
We categorize the keyword mapping task as \textit{relation matching} and \textit{non-relation matching}. The former mapping refers to matching for table or column names, and the latter refers to matching for database values.

\begin{itemize}[leftmargin=*,topsep=0pt]
    \item[-] \textbf{tf-idf:} Similar to ATHENA \cite{ATHENA}, for each unique value present in the database, we first create an exact matching index and then perform tf-idf for tokens in the NLQ. In case of matches to multiple columns, the column with the biggest tf value is chosen as matching. To handle multi-word keywords, we use n-grams of tokens up to $n=3$. For relation matching, we used lexical similarity based on the Edit Distance algorithm.
    \item[-] \textbf{NALIR:} \textcolor{black}{NALIR \cite{NALIR} uses WordNet, a lexical database in which synonyms are stored 
    for relation matching. They calculate similarity for tokens present in the NLQ over WordNet, and determine a matching if the similarity is bigger than a manually defined threshold. 
    For non-relation matching, for each token present in the NLQ, 
    it utilizes regex or full-text search queries over each database column whose type is text. In case of matches to multiple columns, the column which returns more rows, as a result, is chosen as matching. For fast retrieval, we limit the number of rows returned from the query to $2000$, as in the implementation of NALIR.}
    \item[-] \textbf{word2vec:} For each unique value present in the data-base, cosine similarity over tokens in the NLQ is applied to find mappings using pre-defined word2vec embeddings. The matching with the highest similarity over a certain threshold is chosen.
    \item[-] \textbf{TaBERT:} TaBert \cite{tabert} is a transformer-based encoder which generates dynamic word representations (unlike word2vec) using database content. The approach also generates column encodings for a given table, which makes it an applicable keyword mapper for non-relation matching by performing cosine similarity over both encodings. For a particular token, matching with the maximum similarity over a certain threshold is chosen.
\end{itemize}

\textbf{Effectiveness Comparison}

For a fair comparison, we do not apply any pre or post-processing over the NL queries or use an external source of knowledge, such as a keyword parser or metadata extractor. Results are shown in Table \ref{tab:unsupervisedScores}. Each pair of scores represents token-wise accuracy for relation and non-relation matching. For TaBERT, we only report for non-relation matching, because the approach is not applicable to relation matching.

\textcolor{black}{DBTagger outperforms unsupervised baselines in each dataset significantly, by up to $31\%$ and $65\%$ compared to best counterpart for relation and non-relation matching, respectively. For relation matching, the results of all approaches are similar to each other except the word2vec method for the mas dataset. The main reason for such poor performance is that the mas dataset has column names such as \textit{venueName} for which word2vec cannot produce word representations, which radically reduces the chances of semantic matching.}

\begin{table}[t]
    \centering
    \caption{Accuracy scores of keyword mappers for relation and non-relation matching}
    \begin{tabular}{lccc}
    \toprule
         & \multicolumn{3}{c}{\textbf{Database}}\\
         \cmidrule(lr){2-4}
         Baseline & imdb & mas & yelp 
         \\
         \midrule
         tf-idf&0.594-0.051&0.734-0.084&0.659-0.557\\
         NALIR&0.574-0.103&0.742-0.476&0.661-0.188\\
         word2vec&0.625-0.093&0.275-0.379&0.677-0.269\\
         TaBERT&NA-0.251&NA-0.094&NA-0.114\\
         \midrule
         DBTagger&0.908-0.861&0.964-0.950&0.947-0.923\\
         \bottomrule
    \end{tabular}
    \label{tab:unsupervisedScores}
\end{table}

\begin{figure*}[t]
     \centering
     \begin{subfigure}[b]{0.48\textwidth}
         \centering
         \includegraphics[height=6cm]{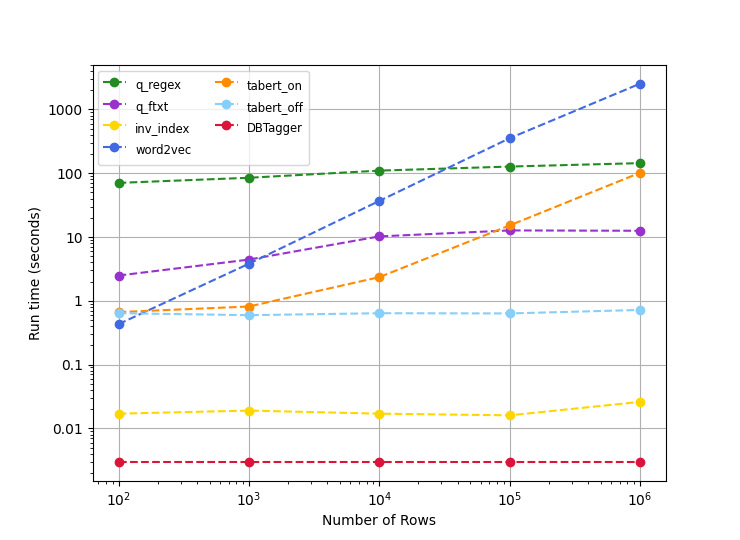}
         \caption{Run Time}
     \end{subfigure}
     \begin{subfigure}[b]{0.44\textwidth}
         \centering
         \includegraphics[height=6cm]{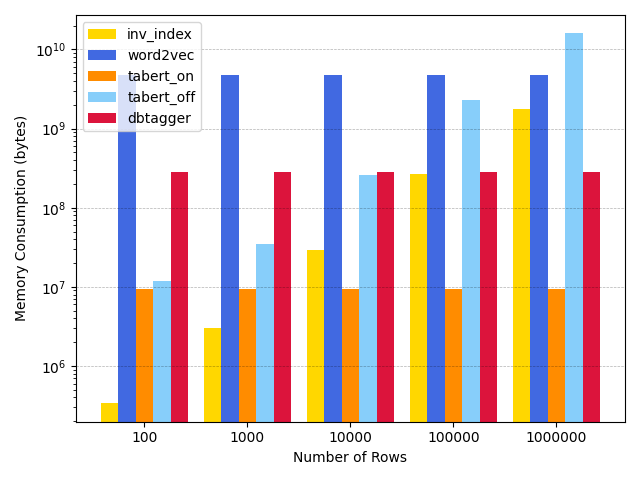}
         \caption{Memory Usage}
     \end{subfigure}
     \vspace{-3mm}
     \caption{\textcolor{black}{Run Time and Memory Usage of state-of-the-art keyword mapping approaches }}
     \label{fig:runtimes}
\end{figure*}

tf-idf gives promising results on the yelp dataset, whereas it fails on the imdb and mas datasets for non-relation matching. This behavior is due to the presence of ambiguous values (the same database value in multiple columns) and not being able to find a match for values having more than three words. For the \textit{imdb} dataset, none of the baselines performs well for non-relation matching. The \textit{imdb} dataset has entity-like values that are comprised of multiple words, such as movie names, which makes it impossible for semantic matching approaches to generate meaningful representations to perform similarity. NALIR's approach of querying over the database has difficulties for the imdb and yelp datasets since the approach does not solve ambiguities without user interaction.


TaBERT performs poorly for all datasets for the non-relation matching task, which we believe is due to two reasons. Firstly, TaBERT has its own tokenizer, which relies on BERT base. The tokenizer tries to deal with out-of-vocabulary tokens by breaking the token into sub-words that have representations. This approach might be useful for a language model; however, it is problematic in the keyword mapping setup since the values present in the databases are domain-specific, which are likely to not occur in the general corpus data used to train such transformers. Also, databases such as imdb, have many entity-like values such as \textit{Eternal Sunshine of the Spotless Mind} which is comprised of several words. Such keywords appearing in the natural language query are therefore divided by the tokenizer into pieces, which eventually leads to unrelated word representations and, thus, non-predictive similarity calculation. The other limitation of TaBERT is its requirement of using cosine similarity. Such an approach requires a manually defined threshold which is not easy to come up with. When a smaller similarity threshold is picked, chances of findind a true positive increases; however, the model becomes prone to generate false positives as well for keywords that are not related to database elements such as stop words and sql specific words (e.g., the, return, find, minimum).

We argue that unsupervised baselines may perform reasonably for relation-matching, whereas they fail to answer the challenges raised by non-relation matching. This is due to the ambiguity present in the databases, such as having values that occur in multiple tables (e.g., "Matt Damon" may appear in both actor and director tables) and domain-specific values that are not covered in word embeddings (e.g., word2vec and TaBERT) trained on general corpus data.

\textbf{Efficiency Comparison}

\textcolor{black}{Efficiency is one of the most important properties of a good keyword mapper to have to be deployable in online interfaces. Therefore, the run-time performance of keyword mapping approaches mentioned in Section 7.3.1 is also evaluated.}

\vspace{-3.5mm}
\textcolor{black}{
\begin{itemize}[leftmargin=*,topsep=0pt]
    \item[-] \textbf{NALIR}: We analyze both querying over database column approaches used in NALIR \cite{NALIR}, named as \textit{q\_regex} and \textit{q\_ftext}, which use \textit{like} and \textit{match against} operators respectively. NALIR \cite{NALIR} uses \textit{q\_regex} approach for tables having less than $2000$ rows and \textit{q\_ftext} for tables having more rows.
    \item[-] \textbf{tf-idf}: Similar to the indexing strategy exercised in AT-HENA \cite{ATHENA}, we implemented an exact matching strategy, using an inverted index named as \textit{inv\_index}, beforehand to avoid querying over the database. The inverted index stores each unique value present in the database along with its frequency in each candidate collection (i.e., database columns).
    \item[-] \textbf{word2vec}: Many works such as Sqlizer \cite{Sqlizer} make use of pre-trained word embeddings to find mappings, which requires keeping the model in the memory to perform the task using cosine similarity.
    \item[-] \textbf{tabert\_on}: TaBert \cite{tabert} requires database content (referred to as content snapshot in the paper) to generate encodings for both NL tokens and columns. We call this setup tabert online, where the model generates the content snapshot on the fly, hence online, to perform mapping when the query comes. 
    \item[-] \textbf{tabert\_off}: We also use TaBert in offline setup. For each table, database content is generated beforehand to perform encodings. In this setup, we keep the content in the memory to serve the query faster.
\end{itemize}
}

\textcolor{black}{We measured the time elapsed for a single query to extract tags and the memory consumption needed to perform mapping for each approach. We also run each experiment with a different number of row values to capture the impact of the database size. Figure \ref{fig:runtimes} presents run time and memory usage analysis of keyword mappers. DBTagger outputs the tags faster than any other baseline, and it is scalable to much bigger databases. However, q\_regex, q\_ftext, tabert\_on, and word2vec do not seem applicable for bigger tables having more than $10000$ rows. The tf-idf technique has a nice balance between run-time and memory usage, but it is limited in terms of effectiveness (Table \ref{tab:unsupervisedScores}). tabert-off performs the tagging in a reasonable time, yet it requires huge memory consumption, especially for bigger tables, and its effectiveness as a candidate keyword mapper is not sufficient. 
}

\begin{table}[t]
\centering
\caption{Overall SQL Query Translation Results}
\begin{tabular}{@{}cccl@{}}
\toprule
\multicolumn{4}{c}{Accuracy (\%)} \\ \midrule
              & xDBTagger       & NALIR+      & NALIR      \\ \midrule
imdb          & \textbf{61.83}      & 50.00       & 38.30      \\
scholar       & \textbf{58.96}      & 40.20       & 33.00      \\
yelp          & \textbf{69.53}      & 52.80       & 47.20      \\ \bottomrule
\end{tabular}
\label{tab:overall-query-results}
\end{table}

\begin{table*}[t]
\centering
\caption{Translation Accuracy Results of xDBTagger According to Categorization of the Queries}
\begin{tabular}{@{}ccccccc@{}}
\toprule
        & \multicolumn{5}{c}{Non-nested} & \multirow{3}{*}{Nested} \\ 
        \cmidrule(r){1-6}
        & \multicolumn{3}{c}{Select-Join (No-Aggregation)} &
        \multirow{2}{*}{Having Aggregation} &  \multirow{2}{*}{Overall} & \\ 
        & Single Table & Multiple Table & Overall & & & \\
        \midrule
        imdb  & 88.89(9) & 68.60(86) & 70.53(95) & 63.64(22) & 69.23(117) & 14 \\
        scholar & 100.00(2)& 86.96(69) & 87.32(71) & 43.59(39) & 71.81(110) & 24  \\
        yelp & 60.00(5)& 83.61(61) & 81.81(66)  & 63.64(55) & 73.55(121) & 7 \\ \hline                 
\end{tabular}
\label{tab:categorized-query-results}
\end{table*}

\begin{table}[t]
\centering
\caption{Accuracy of WHERE Conditions of the Queries with Aggregate Operations}
\begin{tabular}{@{}cccc@{}}
\toprule
        & \begin{tabular}[c]{@{}c@{}}Aggregate Queries\\ with Group By\end{tabular} & \begin{tabular}[c]{@{}c@{}}Aggregate Queries\\ without Group By\end{tabular} & Overall \\
        \midrule
imdb    & 100.00 (6)  & 87.50 (16) & 91.27(22)\\
scholar & 88.89 (18) & 90.48 (21) & 89.53(39)\\
yelp    & 63.64 (11) & 79.55 (44) & 76.80(55)\\ \bottomrule
\end{tabular}
\label{tab:structure-results}
\end{table}

\subsubsection{Query Translation Results}

The numbers of the queries for the three datasets we used in our experiments are provided in Table \ref{tab:comprops}. Also, recall that, we applied 6-fold cross-validation (i.e., leaving 1 fold out for test and using the other 5 folds for training the model) to train our keyword mapper. In order to evaluate query translation results, we performed the translation pipeline for each test fold left out from the training model for \textit{yelp} and \textit{imdb}. We used only 1 test fold for \textit{mas} dataset to make the final number of test queries to be similar to each other. We manually evaluated the translated SQL queries, counting as $CORRECT$ if and only if the translated query is the same as the ground truth in terms of SQL semantics and correct in SQL syntax; and $INCORRECT$ otherwise. Hence, we report binary accuracy results of xDBTagger.

In order to evaluate the comparative performance of xDBTagger, we used two different pipeline-based solutions; namely NALIR \cite{NALIR} and TEMPLAR \cite{jagadishBridge} (an enhanced version of the NALIR, referred as NALIR+, utilizing query logs to detect keyword mappings of the tokens in NLQ). The reason why we choose these two baselines is 3-fold. Firstly, both studies reported accuracy results of their translation pipeline for the same set of three datasets we used in our work. Secondly, both are pipeline-based solutions; that is, they are comprised of sub-solutions for each step in the translation pipeline similar to xDBTagger. Lastly, TEMPLAR \cite{jagadishBridge} tries to enhance the translation pipeline of an existing NLIDB solution (e.g., NALIR) by solely focusing on keyword mappings. Similarly, xDBTagger utilizes the keyword mappings output by DBTagger \cite{dbtagger} in the translation pipeline.

The overall translation accuracy results for xDBTagger, along with the two baselines explained above, are provided in Table \ref{tab:overall-query-results}. Accuracy results of the baselines are taken from the TEMPLAR study \cite{jagadishBridge}. xDBTagger outperforms both baselines in all three datasets, up to $78\%$ and $46\%$ compared to NALIR and NALIR+, respectively. Considering efficiency (see Figure \ref{fig:runtimes} for reference) of the keyword mapper utilized in xDBTagger, simplicity of the translation algorithm explained in Section  \ref{sec:SQLExtract} and having fully explainable end-to-end translation pipeline, the accuracy of xDBTagger stands out even more compared to their counterparts.



One of the limitations of the translation pipeline of xDBTagger is that it fails to translate certain types of queries correctly. The translation algorithm is not able to translate the queries requiring nested SQL queries. The results provided in Table \ref{tab:overall-query-results} include those queries as well; hence less accuracy is observed overall. Another limitation is that queries requiring aggregate operations are difficult to translate (i.e., prone to mistranslation) for xDBTagger
Although we implemented heuristics (see Section \ref{subsec:aggHeuristic} for reference) to address these queries, most of the incorrectly translated queries fall under this category.

In order to further show the efficacy of xDBTagger for other types of queries, we categorized the queries reflecting their difficulty in terms of translation and report accuracy for each category. The results are presented in Table \ref{tab:categorized-query-results}. The numbers in parenthesis represent the number of queries falling under that particular category. The category \textit{Nested} represents the number of queries xDBTagger could not translate due to the translation requiring a nested SQL query. Most of the queries fall under the category \textit{Select-Join with Multiple Tables}, where xDBTagger performs most competitively across all categories. Accuracy results in Table \ref{tab:categorized-query-results} also indicate that heuristics for aggregation queries are effective at translating more than half of the queries under that category on the average.

We further manually evaluated $WHERE$ conditions of the queries having aggregate operations and reported accuracies in Table \ref{tab:structure-results}. As it can be seen, xDBTagger is able to extract $WHERE$ conditions fairly well with 90\% average accuracy for imdb and scholar, and 76\% accuracy for yelp. Although xDBTagger performs the worst for queries having aggregate operations in terms of full translations (Table \ref{tab:categorized-query-results}), it still extracts correct $WHERE$ conditions for both utterances found in the NLQ and the join-path, which is reported to be the most challenging part of the translation in \cite{xu2017sqlnet}. Results show that extracting correct $WHERE$ conditions for the translation is one of the main strengths of xDBTagger.

\subsection{Explainable User Interface of xDBTagger}
\label{sec:user-interface}

\begin{figure*}[!t]
  \includegraphics[width=1.0\textwidth]{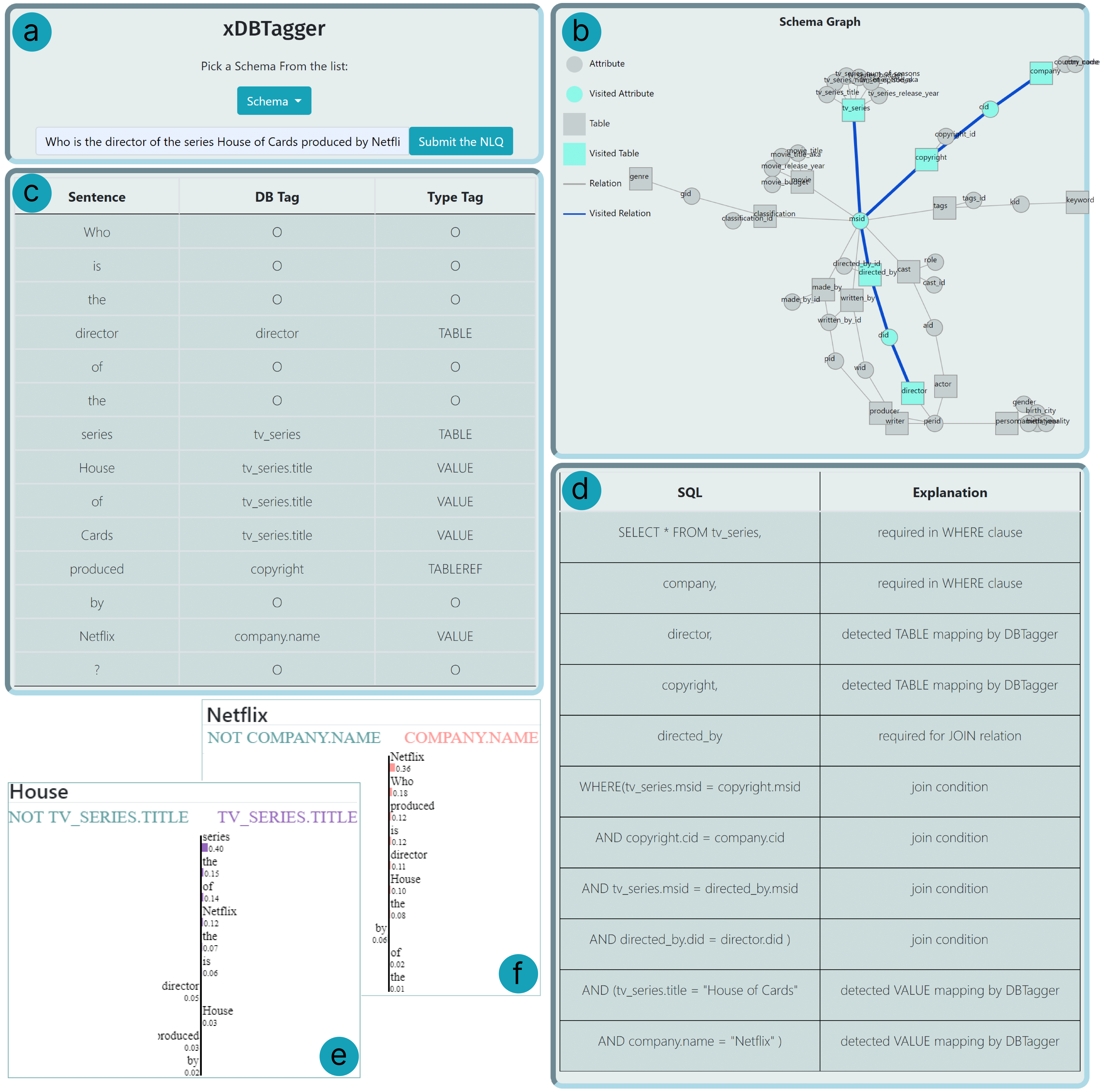}
\caption{Interface components of xDBTagger; (a) NLQ input panel, (b) schema graph panel on which required join path is drawn, (c) predictions panel for keyword mapper, (d) result panel containing the translated SQL query along with its explanations, (e,f) explanation pop-ups for each token in NLQ. Components b,c,d,e and f contain visual and/or textual explanations for the algorithms utilized through the translation pipeline.}
\label{fig:nlidbinterface}       
\end{figure*}

We constructed a simple, single-page web application where users can input a natural language query into our NLIDB pipeline and retrieve the translated SQL. This web application is developed using flask micro-framework and  javascript. Figure \ref{fig:nlidbinterface} depicts interface components through the different stages of the translation pipeline. The pipeline has the following three main steps that we explain to the user separately; 

\begin{enumerate}
    \item Finding type and schema mappings of the tokens in the NLQ using DBTagger. As discussed in Section \ref{sec:keywordMapper}, DBTagger is a sequence-tagging deep learning model, which behaves as a black-box mechanism, i.e., it only classifies the tokens in the NLQ to the most probable class among the possible candidates based on probabilities it derives, yet it does not provide why it makes a particular decision. Therefore, we need to explain how DBTagger maps tokens in the NLQ to the schema elements in the database. To this end, we provide two different interface components. First, we list all the type and schema mappings output by DBTagger for each token in a table (\ref{fig:nlidbinterface}c) to make the decisions made by DBTagger transparent. Second, we visually explain why DBTagger came up with a particular pair of type and schema mapping for a token in a pop-up (\ref{fig:nlidbinterface}e,f) using LIME wrapper (explained in Section~\ref{sec:lime-discussion}) to make it easier for the user to comprehend the decisions made by DBTagger.
    \item Extracting the join path necessary to access tables that include utterances found in the previous step. We first visually draw the schema graph on the panel (depicted in Figure~\ref{fig:nlidbinterface}b) so that the user can better understand the schema underlying the database and the relational connections it inherits. Moreover, we also highlight the nodes and the edges visited along the path on the graph to construct the JOIN clause to explain how certain tables that are not present in predictions panel (\ref{fig:nlidbinterface}c) (i.e., intermediate tables completing the join) appear in the final SQL translation. 
    \item constructing the SQL by forming the WHERE clauses and applying post-processing heuristics to handle certain group of queries. After finalizing the SQL translation, we part-by-part explain how the translated SQL is composed in the result panel, shown in \ref{fig:nlidbinterface}d. In particular, we explain why we include each table in the FROM clause and each logical expression in the WHERE clause. 
\end{enumerate}


NLQ input panel (Figure \ref{fig:nlidbinterface}a) allows the user to select a database schema from a dropdown list and input a query to the NLIDB pipeline for the selected schema.  
After selecting the schema over which the query is to be issued, the schema graph panel displays the extracted graph (i.e., the graph without the highlighted nodes and edges, depicted in Figure \ref{fig:labelledjoinpath}), as explained in Section~5.1. When a query is processed in the pipeline and a SQL is generated for the translation, we highlight the nodes and edges of the graph in blue color, as shown in Figure \ref{fig:nlidbinterface}b, that are used in any part of the generated SQL; (i) tables and their attributes required for the WHERE clauses and (ii) tables required for the correct join operation along with their attributes making the primary-foreign key connections.    



In the predictions panel shown in Figure~\ref{fig:nlidbinterface}c, we present the type and schema mapping outputs of our keyword mapper inside a table so that the user can visually see how his/her query is predicted by the keyword mapper. Furthermore, we used pop-ups to display LIME explanations of each word of the query that is not tagged as 'O' to make it easier for the user to comprehend the decision-making behind the keyword mapper model. When the user clicks on a particular row of the table in Figure~\ref{fig:nlidbinterface}c, a pop-up (e.g., Figure~\ref{fig:nlidbinterface}e) is displayed containing the explanation for the word visually by highlighting the neighboring words that contribute the most, either positively or negatively, when predicting the output.

For example, for the query shown in the input panel Figure~\ref{fig:nlidbinterface}a, we provided the explanation pop-ups for tokens "House" and "Netflix" in Figure~\ref{fig:nlidbinterface}e and Figure~\ref{fig:nlidbinterface}f, respectively. In Figure~\ref{fig:nlidbinterface}e, there are two labels named 'NOT TV\_SERIES.TITLE' and 'TV\_SERIES.TITLE', which correspond to categories of tokens contributing to the label of 'tv\_series.title' negatively and positively, respectively. Below both labels, on each side, there are tokens from the NLQ with a contribution score associated with them. A positive contribution means that the token increases the prediction probability of the explained token for the given class (i.e., TV\_SERIES.TITLE), and a negative contribution decreases that probability. 


For the token in Figure~\ref{fig:nlidbinterface}e, we can see that the word 'series' in the NLQ has the highest positive contribution marginally compared to other tokens. This means that the word 'series' is the most influential neighboring word when determining the mapping classification of the token 'House'. Similarly, Figure \ref{fig:nlidbinterface}f gives the explanation for the word 'Netflix' in the given NLQ. The explanation shows that the word itself has the highest positive contribution, which is expected since the entity is self-expressive and should infer the name attribute of a particular entry in the company table. 




The result panel, shown in Figure~\ref{fig:nlidbinterface}d, presents the generated SQL query and the explanation of how it is composed for the given NLQ. We divide the generated SQL statement into parts and explain why we include each part in the final statement so that users with a less technical background in SQL can better comprehend how the final SQL is composed. In particular, we explain why we include tables and logical expressions in the FROM and WHERE clauses, respectively. For instance, for the example NLQ given in Figure~\ref{fig:nlidbinterface}a, there are three different explanations for why a certain table is included in the FROM clause. 'tv\_series' table is included because there are tokens whose schema mappings are title attribute of the 'tv\_series' table (8-10th rows in the prediction table shown in Figure~\ref{fig:nlidbinterface}c). Director and copyright tables appear thanks to the predictions of DBTagger, whereas 'directed\_by' is present because it is a required table to connect 'tv\_series' and 'director' through join. 

For each logical expression we put in the finalized SQL statement, we provide an explanation as well. Broadly, we divide the explanations into two categories. If a logical expression is to provide a connection between tables in the schema graph, we say it is required for the join condition. If we detect a type mapping of \textit{VALUE} (e.g., 8-10th and 13rd rows in the prediction table shown in Figure~\ref{fig:nlidbinterface}c), we state that a value is detected by DBTagger as shown in the last two rows in the explanation table in Figure~\ref{fig:nlidbinterface}d.

\section{Related Work}
\label{sec:relatedWork}

Although the very first effort \cite{firstNLIDB} of providing natural language interface in databases dates back to multiple decades ago, the popularity of the problem has increased due to some recent pipeline-based systems proposed by the database community, such as SODA \cite{SODA}, NALIR \cite{NALIR}, ATHENA\cite{ATHENA} and SQLizer\cite{Sqlizer}.

However, with the recent advancements in deep neural networks, the problem of NLIDB has also attracted researchers from the NLP community. \cite{zhong2017seq2sql} provided a dataset called \textit{WikiSql} to the research community working on NLIDB problem for evaluation. WikiSql is comprised of $26,531$ tables and $80,654$ pairs which can be used for input for the translation problem. Consequently, many works \cite{zhong2017seq2sql, xu2017sqlnet, yavuz-etal-2018-takes, huang-etal-2018-natural, yu-etal-2018-typesql} utilizing encoder-decoder abstraction have been proposed to evaluate their translation solutions on the WikiSql dataset. However, the dataset only includes schemas with a single table, limiting detailed evaluation of the solutions due to simplicity. 

To remediate this limitation, \textit{Spider} dataset is provided in the work \cite{yu-etal-2018-spider} to the community. Many studies utilizing the pioneer work BERT \cite{devlin-etal-2019-bert}, a pre-trained language model based on the transformer \cite{vaswaniAttention} architecture, have evaluated their solutions on Spider \cite{yu-etal-2018-spider} dataset. Some works \cite{IRNet2019-towards, lin-etal-2020-bridging} focus on the schema linking process to enrich the input NLQ for better leveraging the schema information. In IRNet \cite{IRNet2019-towards}, the authors first query n-grams of the NLQ over the database elements to find candidates and then feed these found candidates to the additional schema encoder, whereas Lin et al. \cite{lin-etal-2020-bridging} integrate these found candidates into the input as a serialization technique before encoding. 
Rat-SQL\cite{wang-etal-2020-rat} proposes a modified transformer layer to leverage schema information better by introducing bias towards the schema for the attention mechanism. 
In addition to these studies, language representation techniques utilizing BERT such as TaBERT \cite{tabert} and Grappa \cite{deng-etal-2021-Grappa} have been introduced to leverage tabular data specific representations in related downstream tasks such as NLIDB problem. For a comprehensive survey covering existing solutions in NLIDB, the reader can refer to \cite{survey2019, survey2020}.

To the best of our knowledge, there is no hybrid solution utilizing both neural network and rule-based techniques, proposed similar to xDBTagger. Nonetheless, some of the earlier works \cite{yavuz-etal-2018-takes, yu-etal-2018-typesql, IRNet2019-towards, lin-etal-2020-bridging} embracing end-to-end neural network approaches focused more on enriching the input by trying to map tokens in the NLQ to database values similar to our keyword mapper, DBTagger. However, as shown in the work \cite{dbtagger}, such solutions proposing ad-hoc querying over database tables to find candidate mappings are not efficient and not scalable to bigger databases unlike xDBTagger. 



There have been previous studies \cite{explainStructureQueries, sqlSummarize1, sqlSummarize2, deutch2020explaining} proposed in line with interpretable interfaces to databases. However, such solutions rather focus on providing additional information for the SQL query results in the form of summarized texts or snippets exploiting signals of the tuples returned by the result SQL. In this work, our goal is not to explain query results but to explain the decisions that lead to the result SQL for each step in the translation pipeline. To our knowledge, xDBTagger is the first NLIDB system exercising XAI principles to explain how the translation is performed.

\section{Conclusion}
\label{sec:conclusion}

In this work, we presented xDBTagger, the first end-to-end explainable NLIDB solution to translate NLQs into their counterpart SQLs. xDBTagger is a hybrid solution taking advantage of both deep learning and rule-based approaches. First, we detect keyword mappings of the tokens in the input NLQ using a novel deep learning model trained in a multi-task learning setup. Next, we explain the decisions for the keyword mappings using a modified version of a state-of-the-art XAI solution LIME \cite{lime}. We visually illustrate the importance of each surrounding word for each mapping by highlighting their contributions which can be either positive or negative. In addition, we draw the schema graph to visualize better the database schema over which the query is issued. We also color the nodes representing tables and attributes in the graph to explain how the required join conditions in the result SQL are extracted. Finally, we explain each part of the result SQL to the user by providing the reason why we need that particular part given the input NLQ. Our quantitative experimental results indicate that in addition to being fully explainable, xDBTagger is effective in terms of translation accuracy and more preferable compared to other pipeline-based solutions in terms of efficiency.  

\begin{acknowledgements}
This research is supported by The Scientific and Technological Research Council of T\"{u}rkiye \hfill \break (T\"{U}B\.{I}TAK) under the grant no 118E724.
\end{acknowledgements}

%
%

\bibliographystyle{spbasic}      
\bibliography{references}   
\end{document}